\documentclass[runninghead]{llncs}
\pdfoutput=1
\usepackage[T1]{fontenc}

\usepackage{algorithm}
\usepackage{algpseudocode}
\usepackage{comment}
\usepackage{graphicx}
\usepackage[colorlinks,urlcolor=black,linkcolor=blue,citecolor=blue]{hyperref}
\usepackage{lipsum}
\usepackage{listings}
\usepackage{MnSymbol}  \usepackage{siunitx}
\usepackage{todonotes}
\usepackage{xcolor}
\usepackage{xspace}
\usepackage{cleveref}  

\newcommand{\SWH}{Software Heritage\xspace}
\newcommand{\SWHID}[1]{\href{https://archive.softwareheritage.org/#1}{\small\ttfamily #1}}

\newcommand{\SWHSCANNER}{\texttt{swh-scanner}\xspace}
\newcommand{\KB}{\textit{kb}\xspace}
\newcommand{\KNOWN}{\textit{known}\xspace}
\newcommand{\UNKNOWN}{\textit{unknown}\xspace}
\newcommand{\KNOWNPARTITION}{\ensuremath{\KNOWN \cupdot \UNKNOWN}\xspace}

 \newcommand{\SWHfilesApprox}{12 billions\xspace}
\newcommand{\SWHcommitsApprox}{2.5 billions\xspace}
\newcommand{\SWHprojectsApprox}{180 million\xspace}

\newcommand{\DataRepos}{\num{16 845}\xspace}
\newcommand{\DataMaxRepoSizeApprox}{2 million\xspace}
\newcommand{\DataMaxPctFastScan}{95\%\xspace}
\newcommand{\DataMaxPctOkScan}{99\%\xspace}
\newcommand{\DataFastScanTimeApprox}{1 second\xspace}
\newcommand{\DataOkScanTimeApprox}{4.9 seconds\xspace}
\newcommand{\DataMeanScanTime}{0.34 seconds\xspace}
\newcommand{\DataMeanSwhidRatio}{15.4\%\xspace}
\newcommand{\DataMedianSwhidRatio}{2.3\%\xspace}

\begin{document}

\title{Efficient Prior Publication Identification\\ for Open Source Code}
\author{Daniele Serafini\inst{1} \and Stefano Zacchiroli\inst{2}\orcidID{0000-0002-4576-136X}}

\institute{University of Turin, Italy\\ \email{me@danieleserafini.eu} \and LTCI, Télécom Paris, Institut Polytechnique de Paris, France\\ \email{stefano.zacchiroli@telecom-paris.fr}}

\maketitle
\begin{abstract}
  Free/Open Source Software (FOSS) enables large-scale reuse of preexisting
  software components. The main drawback is increased complexity in software
  supply chain management. A common approach to tame such complexity is
  \emph{automated open source compliance}, which consists in automating the
  verification of adherence to various open source management best practices
  about license obligation fulfillment, vulnerability tracking, software
  composition analysis, and nearby concerns.

  We consider the problem of auditing a source code base to determine which of
  its parts have been published before, which is an important building block of
  automated open source compliance toolchains.  Indeed, if source code
  allegedly developed in house is recognized as having been previously
  published elsewhere, alerts should be raised to investigate where it comes
  from and whether this entails that additional obligations shall be fulfilled
  before product shipment.

  We propose an efficient approach for prior publication identification that
  relies on a knowledge base of known source code artifacts linked together in
  a global Merkle direct acyclic graph and a dedicated discovery protocol. We
  introduce \SWHSCANNER, a source code scanner that realizes the proposed
  approach in practice using as knowledge base \SWH, the largest public archive
  of source code artifacts. We validate experimentally the proposed approach,
  showing its efficiency in both abstract (number of queries) and concrete
  terms (wall-clock time), performing benchmarks on \DataRepos real-world
  public code bases of various sizes, from small to very large.
\end{abstract}


\section{Introduction}
\label{sec:intro}

Free/Open Source Software (FOSS) guarantees, among other fundamental user
freedoms, the ability to build upon existing FOSS components when creating new
software applications of any kind. After several decades of constant growth
this has led to present-day massive reuse of FOSS components. A recent
analysis~\cite{synopsis2020ossra} by a major industry player in the field of
mergers and acquisitions (M\&A) software audits reports that 99\% of code bases
audited in 2019 contained open source software components, with 70\% of all
audited code being itself open source.

``But there ain't no such thing as a free [software] lunch'', as the saying
goes. To fully reap the benefits of massive open source software reuse, the
integration of FOSS components into enterprise development processes requires
proper management of the (open source) software supply
chain~\cite{harutyunyan2020osssupplychain}. In particular, attention should be
devoted to component selection and
validation~\cite{spinellis2019ossselection,mcaffer2019ossgovernance}, licence
obligation fulfillment~\cite{german2012oslc}, and tracking known security
vulnerabilities in reused components. Most of these concerns are in fact
\emph{not} specific to open source software, but the extent to which FOSS
enables software reuse makes most development teams face them first and
foremost when dealing with open source software.

As a whole, these concerns are referred to as \emph{Open Source Compliance
  (OSC)}, which consists in adhering to all obligations and best practices for
the proper management of FOSS components. Note that open source \emph{license}
compliance (or OSLC~\cite{german2012oslc}) is just a part of OSC, albeit an
often-discussed one due to the variety and complexity of software
licensing~\cite{lindberg2008osslicensing,almeida2017osslicenses}. The
state-of-the-art industry approach for managing the complexity of OSC---known
as \emph{continuous open source
  compliance}~\cite{phipps2020continuouscompliance}---is to automate as much as
possible the verification of adherence to all obligations and best practices
for FOSS component management and integrate them into continuous integration
(CI) toolchains~\cite{meyer2014ci}.

\emph{Source code scanners} play a fundamental role in open source compliance
toolchains. They are run on local code bases (during CI builds or otherwise) to
identify which code parts are known FOSS components v.~in-house unpublished
code, determine the applicable licenses for the open source
parts~\cite{german2010ninka,gobeille2008fossology}, break down components into
where they come from (known as Software Composition
Analysis~\cite{ombredanne2020sca}, or SCA), produce software bills of
materials~\cite{stewart2010spdx,gandhi2018spdx} (or SBOMs), and automatically
detect license incompatibilities~\cite{kapitsaki2017spdx} or violations of
other best practices (optionally making CI build fails).

In this paper we focus on the first among these problems: \emph{prior
  publication identification}. Given a local source code base to audit, we aim
to efficiently identify which parts of it have been published before, according
to a reference knowledge base of previously published code. Slightly more
formally: we aim to partition the audited source code artifacts into two
non-overlapping sets \KNOWNPARTITION. The \KNOWN partition contains code that
is, according to the reference knowledge base, known to have been published
before (possibly, but not necessarily, under a FOSS license); whereas the
\UNKNOWN partition contains code that is supposed to have been written
in-house, and hence never published before.  Determining prior publication
efficiently is of practical importance because it helps continuous OSC
toolchains to ``fail fast''~\cite{gray1986failfast}. Indeed, if one can quickly
determine that supposedly in-house code has in fact been published before, that
is often reason enough to raise an alert, e.g., by making a CI build fail. That
will in turn trigger further investigation, usually by Open Source Program
Office~\cite{mcaffer2019ossgovernance} (OSPO) staff, to determine where the
unknowingly reused code comes from and whether additional overlooked
obligations (legal or policy) concerning it need to be fulfilled before product
shipment.

\paragraph{Contributions.}

With this paper we contribute to improve the state-of-the-art of open source
compliance toolchains for large software systems as follows:
\begin{itemize}

\item We propose an efficient approach for prior publication identification
  based on: (1) a source scanner running locally on the code base under audit;
  (2) a (remote or local) knowledge base of known source code artifacts,
  indexed as a global Merkle DAG~\cite{Merkle}; (3) a discovery protocol
  between the two, called \emph{layered discovery}, that minimizes the amount
  of artifact identifiers whose known/unknown status has to be queried from the
  knowledge base.

\item We introduce \SWHSCANNER, a novel source code scanner that realizes the
  proposed approach in practice, establishing its feasibility. As its default
  knowledge base \SWHSCANNER uses \SWH~\cite{swhcacm2018}, the largest public
  archive of publicly available source code artifacts, having archived and
  indexed (at the time of writing) \SWHfilesApprox unique source code files and
  \SWHcommitsApprox unique commits from more than \SWHprojectsApprox
  development projects. Alternative knowledge bases can be used instead of
  \SWH, e.g., to better cope with inner
  source~\cite{stol2014innersource,capraro2017innersource} use cases.
  \SWHSCANNER is open source software developed at
  \url{https://forge.softwareheritage.org/source/swh-scanner/} and distributed
  via PyPI under the name \texttt{swh.scanner} (see \Cref{sec:implementation}).

\item We validate experimentally the proposed approach, by analyzing \DataRepos
  real-world public code bases of various sizes, from a handful up to
  \DataMaxRepoSizeApprox source code files and directories. Benchmark results
  show that the proposed approach is efficient in terms of how many source code
  artifacts have to be looked up from the knowledge base w.r.t.~its total size
  (\DataMeanSwhidRatio on average). Benchmarks also show that \SWHSCANNER is
  efficient enough for both interactive use and CI integration.
  \DataMaxPctFastScan of the tested code bases can be scanned in less than
  \DataFastScanTimeApprox using \SWH as knowledge base (\DataMaxPctOkScan in
  less than \DataOkScanTimeApprox), with a mean scan time of \DataMeanScanTime.

\end{itemize}

\noindent
In the context of open source supply chain management, \emph{open
  compliance}~\cite{koltun2011opencompliance,fendt2019opencompliance} refers to
the goal of pursuing compliance by only using open technology, including open
source software, open data information, and open access documentation
(including standard specifications). Open compliance helps with reducing
lock-in risks towards service providers and helps with establishing trust in
the scanning tools when they need to run on sensitive code bases. As a
byproduct of the chosen approach, and when \SWH is used as knowledge base,
\SWHSCANNER is the first open-compliance-compliant source code scanner, for the
specific purpose of prior publication identification.

 \section{Approach}
\label{sec:approach}

The problem we aim to solve can be stated as follows. The scanner takes as
input: (1) a local code base to audit, i.e., a source code tree rooted at a
``root'' directory on the filesystem that (recursively) contains all relevant
source code files, and (2) a knowledge base (``KB'' for short) capable of
answering queries about whether individual source code files or directories are
known to have been previously published or not, based solely on their
content---so that, for instance, once a given version of a \texttt{hello.c}
file has been observed in a given Git repository, it will be considered to be
\KNOWN no matter in how many different directories or repositories it appears
in the future.

The scanner produces as output a \KNOWNPARTITION partition of the input code
base, where each audited source code file belongs to either the \KNOWN or
\UNKNOWN set. An input file will belong to the \KNOWN partition if and only if
it is reported as ``known'' by the KB. Note that, the output being a
\emph{partition} of the input, it also holds that \KNOWNPARTITION is equal to
the full set of scanned files.

A couple of caveats are worth noting. First, the input code base can contain
duplicate files, or even duplicate directories. As the known/unknown
determination by the knowledge base depends only on their content, all
different occurrences of the same file (or directory) in the input source tree
will belong to the same output partition. Second, when all files contained in a
source directory belong to the same partition (say, \KNOWN) we can say as a
shorthand that the directory itself belongs to that partition, but the final
partitioning can always be described as the set of all files contained
recursively in the root directory.

\subsection{Knowledge base}
\label{sec:kb}

Without any additional information about the structure of scanned source code
artifacts, the best one can do to establish the \KNOWNPARTITION partition is to
query the KB for \emph{all} individual files. Doing so can incur significant
costs for large code bases. For example, version 5.9.1 of the Linux kernel
contains \num{327441} files and it is going to be just \emph{a} part of mixed
FOSS/proprietary code bases for IoT devices that use Linux as embedded
operating system. This naive approach of querying the status of all source code
files and directories is our baseline of (non) efficient prior publication
identification.

The centerpiece of the proposed approach is a Merkle~\cite{Merkle} DAG (Direct
Acyclic Graph) that links together all source code artifacts known to the KB.
In Merkle structures node labels are not chosen, but computed as strong
cryptographic identifiers based only on the content of each node and, for
non-leaf nodes, on the identifiers of their children. State-of-the-art
distributed version control systems (DVCSs) such as Git~\cite{git} already rely
on Merkle structures, as do P2P filesystems like IPFS~\cite{benet2014ipfs} and
Distributed Ledger Technologies (DLTs).

Merkle structures enjoy properties which are useful for efficient comparison of
structured data. In particular it holds that if a \emph{complete} Merkle
structure (i.e., one with no outgoing dangling links from any node in it)
contains a given node, then it also contains all of its descendants, leaves or
otherwise.

\begin{figure}[tb]
  \centering
\includegraphics[width=0.9\textwidth]{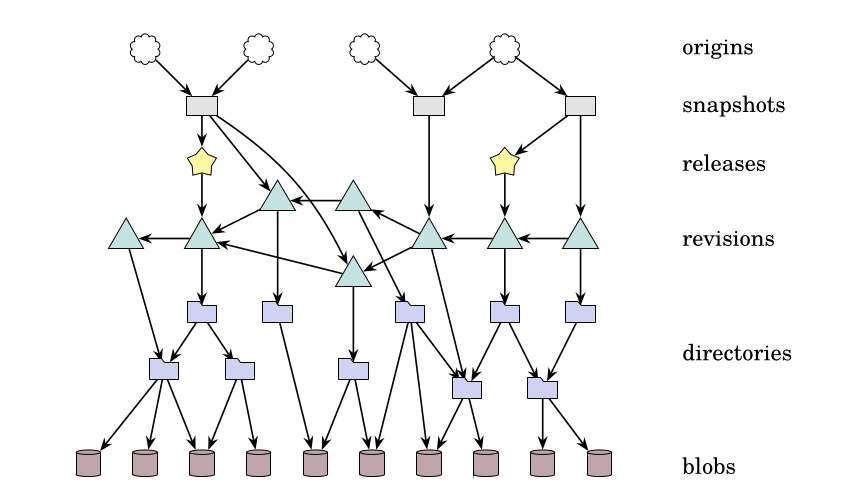}
  \caption{Data model of the global corpus of public software development: a
    Merkle DAG (Direct Acyclic Graph) linking together deduplicated source code
    artifacts.}
  \label{fig:swh-model}
\end{figure}

A generic data model for representing source code artifacts commonly stored in
VCSs has been introduced by \SWH (SWH)~\cite{swhipres2017} for long-term
archival needs; it is shown in \Cref{fig:swh-model}. The SWH data model
supports individual source code files (or ``\emph{blobs}'' in SWH jargon),
\emph{directories} (i.e., source code trees), commits (or
``\emph{revisions}''), \emph{releases} (i.e., commits annotated with mnemonic
labels like ``2.0''), and repository \emph{snapshots} (i.e., the full state of
repositories, keeping track of where each branch was pointing at archival
time). \emph{Origin} nodes represent software distribution places, such as
public Git repositories identifier by URLs, and act as the graph roots pointing
\emph{into} the Merkle DAG.

As node identifiers \SWH, and by extension \SWHSCANNER, relies on \emph{SWHIDs}
(SoftWare Heritage IDentifiers)~\cite{swhipres2018}, which are standardized
intrinsic textual identifiers that embed a cryptographic checksum (a SHA1, in
SWHID version 1) and node type information. Examples of a blob and directory
SWHID identifier are:
\SWHID{swh:1:cnt:94a9ed024d3859793618152ea559a168bbcbb5e2} and\linebreak
\SWHID{swh:1:dir:d198bc9d7a6bcf6db04f476d29314f157507d505}. SWHIDs can be
resolved via the SWH archive Web UI and various other public
resolvers.\footnote{For details see
  \url{https://docs.softwareheritage.org/devel/swh-model/persistent-identifiers.html}.}

We require the KB to index source code artifacts using a Merkle DAG, to support
at least directory and blob nodes, and to behave consistently w.r.t.~Merkle
properties when answering known/unknown queries. In particular, if the lookup
of the known status of a directory returns \KNOWN, then the knowledge base must
also return \KNOWN for all source code files and directories (recursively)
contained in it.  On the other hand the proposed approach does not
\emph{depend} on SWH specifically. We have built \SWHSCANNER (see
\Cref{sec:implementation}) using the \SWH archive due to its availability and
large coverage of public code, but the proposed approach is applicable to any
KB respecting the desired Merkle properties.  Node types other than files and
directories, while not strictly needed for code tree scanning, can also be
exploited if available (see \Cref{sec:discussion}).

\subsection{Discovery protocol}
\label{sec:algo}

\begin{algorithm}[tb]
  \caption{\emph{Layered discovery}: efficient prior publication identification
    of known source code artifacts}
  \label{algo:layered}
  \begin{algorithmic}[1]
  \Procedure{LayeredDiscovery}{$\textit{root}, \KB$}
    \State $\KNOWN, \UNKNOWN \gets \emptyset, \emptyset$
      \Comment{output partition}
    \State $Q \gets \emptyset$ \Comment{queue of nodes to visit}
    \State $Q.\textit{enqueue}(\textit{root})$
    \While{$Q \neq \emptyset$} \Comment{main loop}
      \State $\textit{node} \gets Q.\textit{dequeue}()$
      \If{$\KB.\textit{knows}(\textit{node}.\textit{id})$}
          \Comment{knowledge base lookup}
        \State $\KNOWN \gets \KNOWN \cup \{\textit{node}\}$
        \If{$\textit{node}.\textit{type} = \textit{directory}$}
          \State \Comment{known directory, mark descendants as known}
          \State $\KNOWN \gets \KNOWN \cup \textit{visit}(\textit{node})$
        \EndIf
      \Else \Comment{unknown node}
        \State $\UNKNOWN \gets \UNKNOWN \cup \{\textit{node}\}$
        \If{$\textit{node}.\textit{type} = \textit{directory}$}
          \State \Comment{unknown directory, dig further}
          \For{$\textit{child} \in \textit{children}(node)$}
            \State $Q.\textit{enqueue}(\textit{child})$
          \EndFor
        \EndIf
      \EndIf
    \EndWhile
    \State \Return $\langle\KNOWN,\UNKNOWN\rangle$
      \Comment{return partition}
  \EndProcedure
  \end{algorithmic}
\end{algorithm}

Once a knowledge base is available, a source code scanner can efficiently
determine the \KNOWNPARTITION using the \emph{layered discovery} protocol
detailed in \Cref{algo:layered}. It takes as first input \textit{root}, the
root node of a Merkle tree that corresponds to the code base to audit, indexed
in the same way used by the knowledge base to index known code\footnote{In
  particular, Merkle node identifiers shall be computed in the exact same way
  between the knowledge base and the scanner.} and containing directory and
blob nodes. Note that, due to the fact that Merkle structures are built
bottom-up, the scanner needs to read the entire local code base to obtain the
root node before starting \Cref{algo:layered}. The second input is \KB, a
handle to the knowledge base that can be queried to obtain the known/unknown
status of a given node identifier.

The following additional notation is used in \Cref{algo:layered}: nodes have
two attributes, $n.\textit{type}$ returning the node type (blob or directory)
and $n.\textit{id}$ returning its Merkle identifier; $\textit{visit}(n)$ is a
function returning all the (recursive) descendants of node $n$, including
itself, in an arbitrary order; $\textit{children}(n)$ returns the direct
(non-recursive) descendants of a given node; $Q$ is a FIFO queue, equipped with
the usual $Q.\textit{enqueue}(n)$ and $Q.\textit{dequeue}()$ methods. The
knowledge base is equipped with a single $\KB.\textit{knows}(\textit{id})$
method that returns a boolean indicating whether it knows about a given node or
not, based on its Merkle identifier.\footnote{Hash collisions are possible, but
  we assume that the chosen cryptographic hash function is strong enough for
  the target domain.}

Layered discovery proceeds by performing a BFS (breadth-first search) visit of
the local source tree, querying the knowledge base and updating the (initially
empty) \KNOWN/\UNKNOWN partition as it goes. Once a known node is encountered,
knowledge base querying can stop, because all nodes in its subgraph must be
known to the knowledge base as well, due to Merkle properties. This allows to
prune potentially large subgraphs, minimizing querying, which is expected to be
costly, as it will usually happen over the network. However, note that when
querying stops, the entire subgraph should still be added to the \KNOWN
partition; this step can be done locally without further interacting with the
knowledge base.

\paragraph{Runtime complexity.}

In the worst case scenario, when \KB does not know any node of the local code
base, layered discovery has a time-complexity of $2 \cdot O(V + E) = O(V + E)$
(one bottom-up visit of the source tree to build the Merkle DAG plus one
top-down visit to determine the \KNOWNPARTITION split), where $V$ and $E$ are
respectively the nodes and edges of the audited code base as a Merkle DAG; the
scanner will perform $O(V)$ knowledge base lookups using
$\KB.\textit{knows}()$.

In the best case scenario, when \KB knows the entire code base, complexity is
$O(V + E)$ with a single call ($O(1)$) to $\KB.\textit{knows}(\textit{root})$.

We will verify experimentally in \Cref{sec:validation} that, using \SWH as a
knowledge base, we are often close to the best case scenario and also that,
independently from the chosen knowledge base, this approach is significantly
more efficient than the baseline. We will also see that runtime in practice is
good enough for both interactive and CI integration use cases. Before that, let
us see how \SWHSCANNER implements in practice the proposed approach.

 \section{Design and implementation}
\label{sec:implementation}

\begin{figure}[tb]
  \centering
  \includegraphics[width=\textwidth]{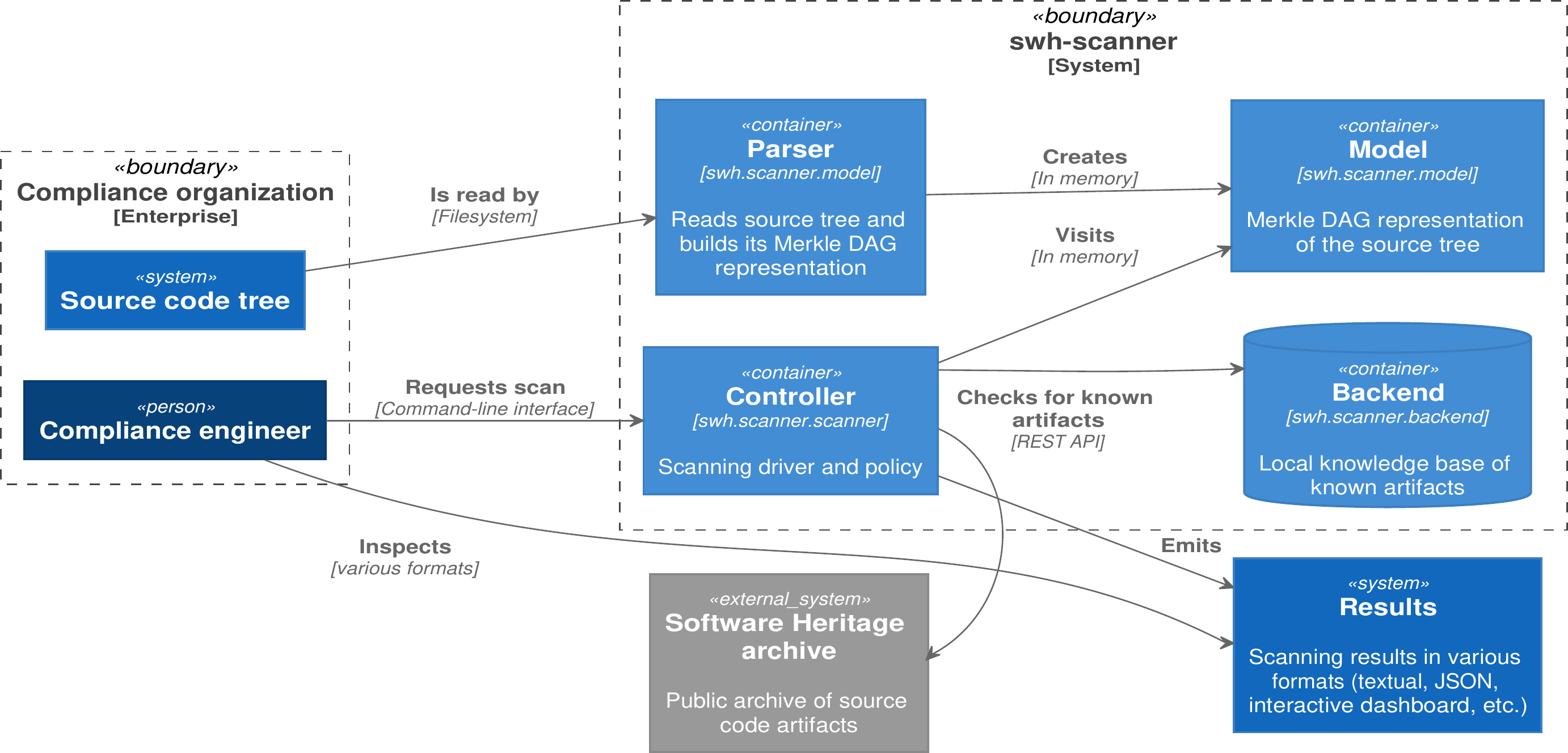}
  \caption{Architecture of \SWHSCANNER as a C4 container diagram.}
  \label{fig:architecture}
\end{figure}

\Cref{fig:architecture} shows the architecture of \SWHSCANNER as a C4 container
diagram. In this section we describe the role of each component in the
architecture, throughout the execution of a typical \SWHSCANNER use case.

On the left of \Cref{fig:architecture} a compliance engineer of some
organization is interested in establishing the prior art status of a local
source code tree. \SWHSCANNER, on the top-right of the figure, is free/open
source software, released under the GNU GPL license version 3 or above,
implemented in Python and distributed via PyPI. It can be installed using the
\texttt{pip} package manager as follows:
\begin{lstlisting}[language=bash]
$ pip install swh.scanner
\end{lstlisting}

\noindent
The compliance engineer can then run \SWHSCANNER on the input source tree:
\begin{lstlisting}[language=bash]
$ swh scanner scan SRC_ROOT/
\end{lstlisting}
where \texttt{SRC\_ROOT} is the root directory of the source code tree for
which we want to determine prior art. Upon invocation \SWHSCANNER will first
build, using the \emph{Parser} component, an in-memory \emph{Model} of the
source tree as a Merkle DAG structure compatible with the one described in
\Cref{sec:kb}.

The \emph{Controller} component will then run layered discovery (see
\Cref{algo:layered}) on the source tree model to determine the \KNOWNPARTITION
partition. \SWHSCANNER also implements alternative, user-selectable discovery
protocols (e.g., scan directory nodes first, scan file nodes first, random
scanning, etc.) in addition to layered discovery, which is the default and most
efficient protocol.

By default \SWHSCANNER directs KB queries---corresponding to \emph{kb.knows()}
method invocations in \Cref{algo:layered}---to the \SWH public REST API, which
provides a dedicated \texttt{/known}
endpoint\footnote{\url{https://archive.softwareheritage.org/api/1/known/doc/}}
capable of returning the known/unknown status of several SWHIDs at once.
Alternatively, a local knowledge base of known source code artifacts,
identified by SWHIDs, can be operated locally. In such a scenario \SWHSCANNER
can be pointed to the non-default, local KB via the \texttt{web-api}
configuration setting. The companion command \lstinline[language=bash]|swh scanner db serve -f KB.sqlite| will serve a \SWHSCANNER-compatible KB (on a configurable host/port) using the
\texttt{KB.sqlite} database as source of truth for known SWHIDs. More complex
setups are possible by providing a custom implementation of the \texttt{/known}
endpoint; for instance, in inner source~\cite{stol2014innersource} scenarios
one might want to check first a local KB and then fallback to SWH.

At the end of scanning the Controller emits \emph{Results} in various
user-selectable formats: textual output for manual inspection (\emph{à la}
recursive \texttt{ls}, with annotations of which parts of the tree are
known/unknown), machine-parsable JSON reports, or an interactive HTML dashboard
to drill down into scanning results. For example, a sample run of \SWHSCANNER
on a locally modified version of a Linux kernel source tree, requesting
detailed output in JSON format (e.g, for further automated processing), could
look like this (excerpt):
\begin{lstlisting}[language=bash,basicstyle=\ttfamily\scriptsize]
$ time swh scanner scan -f json /srv/src/linux/kernel
{
  [...]
  "/srv/src/linux/kernel/auditsc.c": {
    "known": true,
    "swhid": "swh:1:cnt:814406a35db163080bbf937524d63690861ff750"
  },
  "/srv/src/linux/kernel/backtracetest.c": {
    "known": true,
    "swhid": "swh:1:cnt:a2a97fa3071b1c7ee6595d61a172f7ccc73ea40b"
  },
  "/srv/src/linux/kernel/bounds.c": {
    "known": true,
    "swhid": "swh:1:cnt:9795d75b09b2323306ad6a058a6350a87a251443"
  },
  "/srv/src/linux/kernel/bpf/": {
    "known": false,
    "swhid": "swh:1:dir:fcd9987804d26274fee1eb6711fac38036ccaee7"
  },
  "/srv/src/linux/kernel/capability.c": {
    "known": true,
    "swhid": "swh:1:cnt:1444f3954d750ba685b9423e94522e0243175f90"
  },
  [...]
}
0,53s user 0,61s system 145$
\end{lstlisting}

 \section{Experimental validation}
\label{sec:validation}

In order to validate the proposed approach for efficient prior publication
identification of open source code artifacts we have used \SWHSCANNER to
analyze \DataRepos public code bases.

We initially selected \num{20 000} public Git repositories from the \SWH
dataset~\cite{swh-msr2019-dataset} whose sizes, measured as the number of
commits in the repository, uniformly distributed on a log scale. This gave us a
varied project sample in terms of project age and activity. We then cloned each
of those projects from their public repository URLs. It is important to notice
that we explicitly did not retrieve the projects from the SWH archive to allow
active projects to have more code and commits w.r.t.~the last archived version
in SWH, e.g., due to archival lag, which is a common scenario in compliance use
cases: the prior art knowledge base used at scan time is not necessarily
up-to-date w.r.t.~the state of public code in the real world.

Clones were performed using \texttt{\small git clone --depth 1} to retrieve
only the most recent commit of each repository, which will constitute one
\emph{code base} to be scanned. We successfully cloned \DataRepos repositories.
The other repositories were no longer available from their original hosting
places and have been ignored.

We then run \SWHSCANNER\footnote{Specifically, we used the \SWHSCANNER version
  identified by SWHID
  \SWHID{swh:1:rev:979d7c803a1478c1e65a6cf8a827c16a746e3aa1}, archived by SWH.}
on each code base keeping track for each invocation of: n.~of \emph{SWHIDs
  looked up} from the KB (equivalently: n.~of \emph{kb.knows(id)} calls in
\Cref{algo:layered}); \emph{elapsed real time} for scanning; \emph{code base
  size} as the total number of files and directories it contains.
(equivalently: n.~of nodes in the Merkle DAG model). Scanned projects have
sizes ranging from small code bases of a few files+directories (mean: 3160,
median: 132) up to medium and very large code bases (90\% percentile: 4198,
95\%: \num{10283}, max: 2.54 millions files+directories).

\begin{figure}\centering
\includegraphics[width=0.9\textwidth]{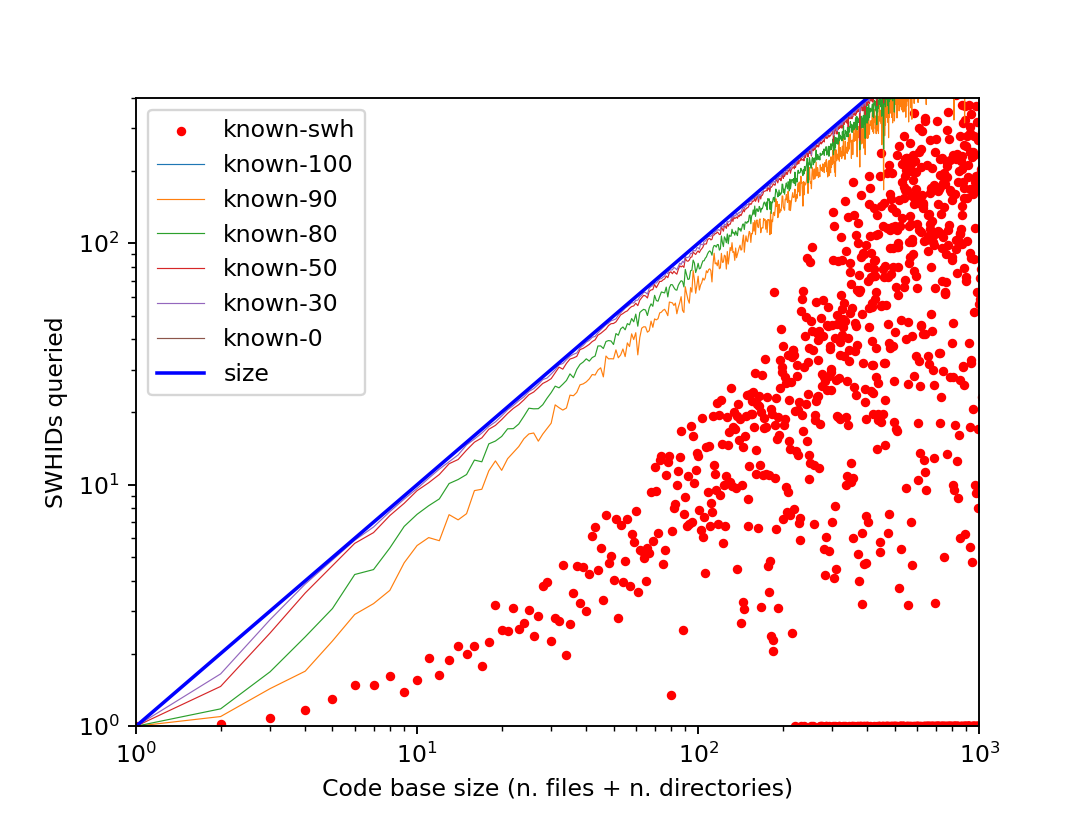}
  \includegraphics[width=0.9\textwidth]{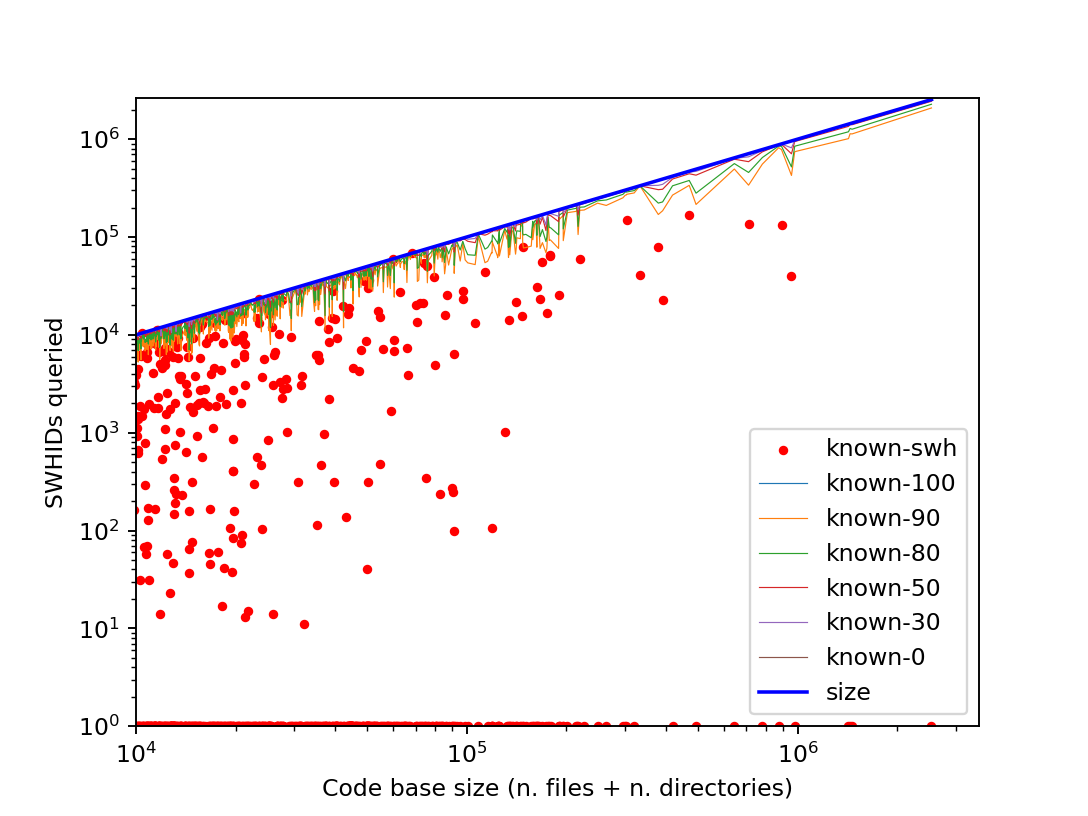}
  \caption{Amount of knowledge base lookups performed for determining the
    \KNOWNPARTITION partition of public code bases of various sizes (small ones
    above, large ones below) using various knowledge bases: \texttt{known-swh}
    for the live \SWH archive, \texttt{known-100} for 100\% of the code base
    files and directories known, down to \texttt{known-10} and \texttt{known-0}
    for 10\% and 0\% artifacts known, respectively.}
  \label{fig:bench-swhids}
\end{figure}

On each code base we have run \SWHSCANNER using different knowledge bases. The
scenario denoted as \texttt{known-swh} consists of using the live \SWH archive
as knowledge base at the time of scanning; it best represents a real compliance
engineer using \SWHSCANNER.  Other scenarios, denoted
\texttt{known-0,known-10,\ldots,known-100}, are simulations of different
knowledge bases knowing from none (0\%, or \texttt{known-0}) to all of (100\%,
or \texttt{known-100}) the files and directories encountered in \emph{all}
tested code bases by 10\% increments.

To obtain the simulated knowledge bases we first mined from all code bases the
identifiers (as SWHIDs) of all contained files and directories. This set
constitutes the \texttt{known-100} knowledge base: any node that could ever be
queried when scanning any code base will be reported as ``known'' by this KB.
To obtain the \texttt{known-90} we proceeded as follows: randomly mark 10\%
files (which correspond to leaves in the Merkle DAG) as unknown; then visit the
Merkle DAG backward from leaves to roots marking all encountered directory
nodes as unknown as well. The visit ensures that Merkle properties are
respected by the simulated KBs: if a given file is reported as unknown, no
directory (recursively) containing it should be reported as known, because if
the directory had been encountered in the wild, all its (recursive) content
would have been too. Iterating this process by 10\% increments we produced all
simulated KBs up to \texttt{known-10}.  (\texttt{known-0} is trivial to
produce: it is the KB that always answers ``unknown''.)

Practically, experiments for the \texttt{known-swh} KB were run using default
scanner settings (which make \SWHSCANNER query the live SWH archive), whereas
other \texttt{known-*} cases by operating local KBs using \texttt{\small swh
  scanner db serve} (discussed in \Cref{sec:implementation}) and pointing the
scanner to them.

\Cref{fig:bench-swhids} shows experimental results about the number of files
and directories looked up from the KB using \emph{kb.knows(id)}, for various
KBs. For the sake of readability results are split between two charts, one for
small code bases (above) and one for big ones (below).  The size line is our
baseline, corresponding to a discovery scenario where the scanner has to lookup
all artifacts (files + directories = size) from the KB. Indeed, the
\texttt{known-0} scenario, i.e., a KB that knows no artifact, is identical to
the baseline (and hidden below it in the charts).

\begin{figure}[tb]
  \centering
  \includegraphics[width=0.9\textwidth]{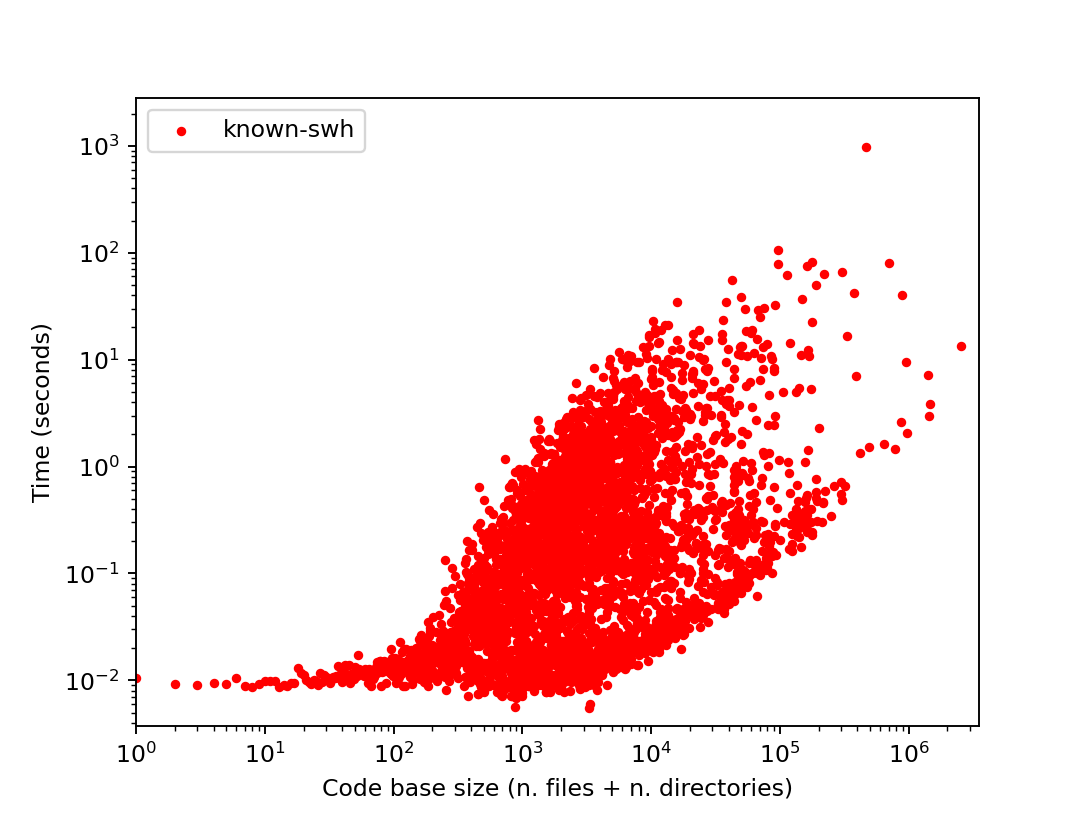}
  \caption{Elapsed real time (seconds) for determining the \KNOWNPARTITION
    partition of public code bases of various sizes using \SWH as remote
    knowledge base. Mean: \DataMeanScanTime; \DataMaxPctFastScan percentile $<$
    \DataFastScanTimeApprox; \DataMaxPctOkScan $<$ \DataOkScanTimeApprox.}
  \label{fig:bench-time}
\end{figure}

The real-world scenario of \texttt{known-swh}, where we have queried the live
SWH archive performs much better than the baseline, looking up only an
increasingly marginal fraction of scanner artifacts. On average as little as
\DataMeanSwhidRatio of the input files and directories need to be looked up,
with a median of \DataMedianSwhidRatio and a 75\% percentile of 20\% nodes of
the input code base looked up.
Among simulated scenarios, the only one outperforming \texttt{known-swh} is
\texttt{known-100}, for a KB that knows all artifacts and hence always need a
single lookup of the root directory node. Other simulated KB scenarios
increasingly approach the baseline: the less the KB knows, the closer they get
to it, i.e., the worse they perform.

These results show that, in terms of lookup efficiency, the proposed approach
beats the baseline and does so by far when using the SWH archive as KB. But
what about the \emph{practical} efficiency and viability of \SWHSCANNER as a
tool for the task? We answer this question by showing in \Cref{fig:bench-time}
the elapsed real time for scanning all analyzed code bases.

Timing benchmarks show that \SWHSCANNER, when used with SWH as remote knowledge
base over the network, is efficient enough for both interactive use, e.g., by a
compliance engineer, and integration into CI/CD workflows for continuous open
source compliance. \DataMaxPctFastScan of the tested code bases can be scanned
in less than \DataFastScanTimeApprox using \SWH as knowledge base
(\DataMaxPctOkScan in less than \DataOkScanTimeApprox), with a mean scan time
of \DataMeanScanTime. Aside from a single outlier (1 code base out of
\DataRepos projects, which took $\approx\,$15 minutes to scan 0.5 million
files/directories) even the largest code bases in our sample, up to 2 million
files/directories, were scanned in less than 2 minutes.

 \section{Related Work}
\label{sec:related}

We have introduced an efficient approach to detect prior publication of open
source code artifacts, implemented it in \SWHSCANNER, and verified
experimentally its efficiency. To the best of our knowledge this is the only
scanner for FOSS compliance that is open source itself, leverages Merkle
properties to improve scanning efficiency and uses \SWH as an open data
knowledge base to determine prior publication of source code files and
directories. While the underlying problem seems to have received little
attention in the research literature a number of industrial code scanning tools
exist, for the purpose of (semi-)automating the verification of compliance with
FOSS license obligations and/or security best practices.

The tooling
landscape\footnote{\url{https://github.com/Open-Source-Compliance/Sharing-creates-value/}}
conducted by the Open Source Tooling Group and the
OpenChain~\cite{coughlan2020openchain}
curriculum\footnote{\url{https://github.com/OpenChain-Project/curriculum/}}
provide a good overview of existing tools to support automated governance of
FOSS supply chains, including tools that adhere to the open compliance
principle~\cite{fendt2019opencompliance} (see \Cref{sec:intro}).
State-of-the-art license scanners in the field are
FOSSology~\cite{jaeger2017fossology}, and ScanCode (discussed
in~\cite{ombredanne2020sca} together with other FOSS tools for Software
Composition Analysis). Zooming out from license detection \emph{per se},
several tools are used in the compliance landscape to manage the workflow of
vetting open source component before production use, such as Eclipse
SW360\footnote{\url{https://www.eclipse.org/sw360/}} as component inventory
manager and the OSS Review Toolkit
(ORT)\footnote{\url{https://github.com/oss-review-toolkit/ort}} that provides a
customizable pipeline for continuous
compliance~\cite{phipps2020continuouscompliance}.

SCANOSS\footnote{\url{https://www.scanoss.com/}} has recently announced an open
data knowledge base (OSSKB) to accompany its (also open) scanning tool. Its
coverage is comparable in size to \SWH, but the indexing technique is
different. OSSKB has finer granularity (see \Cref{sec:discussion} for a
discussion of this point) than \SWHSCANNER, relying on
winnowing~\cite{aiken2003winnowing} for approximate matches on individual
source code files. On the other hand the SCANOSS scanner does not rely on
Merkle structuring to prune code base parts that do not need scanning.
In-depth quantitative benchmarking of the two approaches constitutes
interesting future work.

 \section{Discussion}
\label{sec:discussion}

\paragraph{Scanning commits and other artifact types.}

We focused our discussion on the scanning of source code files and directories,
because that corresponds to both the state-of-the-art in terms of artifact
types and to what \SWHSCANNER supports today. But in fact all artifact types
supported by the SWH data model (\Cref{fig:swh-model})---and in particular
commits, releases and snapshots, not discussed in the paper---can be supported
via the same approach. Minimal changes would be needed in the discovery
protocol, roughly speaking to treat all other non-leaf nodes similarly to how
directories are handled. This flexibility is likely to become increasingly
relevant in the future, as relevant FOSS projects (e.g., the Linux kernel) are
starting to
discuss\footnote{\url{https://lore.kernel.org/linux-spdx/YqILppVZUrD19M6D@ebb.org/}}
the possibility that their \emph{entire Git development history} might
correspond best to the \emph{complete and corresponding source} (CCS) that
should be made available to users for compliance with the terms of the GNU
GPL. In such a scenario it will become important for compliance engineer to
determine the prior publication of entire Git repositories; the proposed
approach will fit well and efficiently such novel use cases.

\paragraph{Scanning granularity.}

Whereas scanning commits, release, etc.~goes in the direction of (conceptually)
larger artifacts, one can also increase scanning granularity and scan
\emph{within} source code files, e.g., to support prior art detection at level
of individual snippets contained in a larger source code file. Various
techniques exist to support this in source code scanners, including plagiarism
detection (like Winnowing~\cite{aiken2003winnowing}, mentioned in
\Cref{sec:related}), locality-sensitive hashing (like
TLSH~\cite{oliver2013tlsh}), or source code parsing followed by code clone
detection~\cite{shobha2021clonedetection}.

While the proposed implementation, based on \SWHSCANNER + SWH as KB, stops at
file granularity with a notion of file equality based on cryptographic
hashes---and hence will \emph{not} recognize as \emph{known} a source code file
where a single byte has been altered w.r.t.~a previously published version of
it---the proposed approach is granularity-agnostic. The Merkle structure of
\Cref{fig:swh-model} can be extended to have as leaves file parts, such as code
snippets or lines (SLOCs), rather than files. That would cause an increase in
the size of the KB, but will not substantially alter the approach efficiency,
because large known sub-parts of public code bases will be fully detected as
known based on hashes that are high up in the Merkle structure. To the best of
our knowledge no attempt of building a Merkle structure of public code at a
granularity finer than individual files and at the scale of SWH (= tens of
billions source code files) has ever been attempted.

\paragraph{Enriching scan results with additional information.}

Industrial source code scanners generally offer as output more information than
mere known/unknown information, as \SWHSCANNER does. The latter not being a
marketed product, we do not consider this a significant limitation: \SWHSCANNER
is meant to be a research prototype showing how the specific sub-problem of
determining prior publication for FOSS artifacts can be solved efficiently
using a Merkle-structured open data knowledge base. At the same time it is
important to discuss how \emph{compatible} the proposed approach is with
``joining'' additional information to scanning output.

Once the \KNOWNPARTITION is identified, computed SWHIDs can be used as unique
keys to lookup additional information about scanned artifacts that can then be
included in scanning results. Typical example of additional information
returned by code scanners for open compliance are: licensing information
(already available from SWH itself, detected using
FOSSology~\cite{jaeger2017fossology}), software composition
analysis~\cite{ombredanne2020sca} decomposition (which would need to be
computed separately), software provenance~\cite{godfrey2015provenance}
information (that can be tracked at the scale of
SWH~\cite{swh-provenance-emse}), and known vulnerability information (available
from public CVE databases, but currently lacking an open data
CVE$\leftrightarrow$SWHID mapping). Once these information become available
from third-party KBs, extending \SWHSCANNER to look them up and join them with
scanning results would be a simple matter of programming.

\subsection{Threats to validity}
\label{sec:threats}

Our experimental benchmarks have been conducted on public code bases rather
than on \emph{private code bases}. We have simulated the ignorance by the
knowledge base of encountered artifacts, but there is no guarantee that matches
the reality of in-house code bases encountered in the real world. It is
difficult to do better while at the same time preserving experiment
reproducibility (on public code that anyone can retrieve and experiment with).
It would nonetheless be interesting to experiment with \SWHSCANNER in an
industrial compliance engineering setting. We are aware of large companies
integrating \SWHSCANNER in their licence compliance toolchains, but no rigorous
empirical experiment has been conducted yet.

The correctness of \SWHSCANNER depends on the fact that the answers returned by
the KB are Merkle-consistent. In particular it must hold that when the KB
answers \emph{known} for a non-leaf node (e.g., a directory), all its
descendants must be \emph{known} as well. In specific corner cases this
property might not hold for the SWH archive. For example, corrupted objects
from some VCS repositories might not have been archived or legal takedown
actions might have forced the archive to poke ``holes'' into the Merkle
structure. Given the Merkle DAG can only be built bottom-op, arguably, holes do
not invalidate the fact that the content that used to be there should be
reported as \emph{known}; after all \emph{it has been observed} in the past,
only to disappear later. A more satisfying answer is technically possible, but
requires engineering a more complex Merkle-based accounting of ``holes''; doing
so is beyond the scope of this paper. In quantitative terms the problem is
negligible when using SWH as KB. It is also a KB-specific issue which does not
impact the validity of the approach as a whole.

 \section{Conclusion}
\label{sec:conclusion}

We introduced an efficient approach to determine prior publication of open
source code artifacts based on a Merkle-structured knowledge base (KB) and
implemented it in the open source \SWHSCANNER tool, which uses the \SWH archive
as KB. By scanning \DataRepos code bases we have experimentally validated the
efficiency of the proposed approach and tool, both in intrinsic terms (calls
between scanner and KB) and in terms of wall-clock time.

Several steps remain as future work. Alternative discovery protocols are
possible: they should be designed, modeled, and benchmarked to determine if
further efficiency improvements are practically viable. Merkle structuring can
also be improved to better cater for real-world data losses on the KB side,
finer artifact granularity, and approximate artifact matching.

 \subsubsection{Acknowledgements}

The authors would like to thank Guillaume Rousseau for providing the project
sample used in the experiments described in \Cref{sec:validation} and, more
generally, for insightful discussions about \SWHSCANNER.

\end{document}